\documentstyle[aps,preprint,amssymb,epsfig]{revtex}

\begin{document}
\tightenlines
\makebox[\textwidth][r]{SNUTP 98-019}

\makebox[\textwidth][r]{UFIFT-HEP-98-8}
\draft

\begin{center}
{\large\bf 
Emission of Axions in strongly
magnetized stars}
\vskip 0.2in

Deog Ki Hong\footnote{
E-mail address: {\tt dkhong@hyowon.cc.pusan.ac.kr}} 
\vskip 0.2in
\it 
Department of Physics, Pusan National 
University, Pusan 609-735, Korea\footnote{permanent address}\protect\\  
and \protect\\
Institute of Fundamental Theory, University of Florida 
Gainesville, Florida 32611, USA
\end{center}

\begin{abstract}
We show that  
the axion decay constant does not get any correction 
at any order 
by external magnetic fields. On the other hand, in 
the context of the Wilsonian effective action under external 
magnetic fields, the axial currents 
get a finite correction.  
We then calculate the effect of strong magnetic fields $B>10^{18}G$ 
on the axion-nucleon coupling and find that 
if $B\gtrsim 10^{20}~G$ in strongly magnetized neutron
and white dwarf stars, the
emission rate of axions is enhanced by 
several orders of magnitude.
\end{abstract}


The stellar evolution tightly constrains the emission of weakly 
interacting, low-mass particles such as neutrinos or 
axions~\cite{raffeltreport} whose properties are otherwise hard to 
probe~\cite{hagman}. 
Recently, the emission rate of such particles from stars 
has been calculated
more accurately by including  
the many-body effects~\cite{seckel},
the effect of strong magnetic fields~\cite{borisov,kachelriess}, 
and the thermal 
pions and kaons~\cite{schramm,ellis}. 
The many-body effects in the dense core of stars  
are found to suppress the axion or neutrino 
emission rate by a factor of about 2, while 
the thermal pions and kaons to   
increase the axion emission rate by a factor of 3-4 within 
a perturbative    
treatment. Under a strong magnetic field $B$
the cyclotron emission of neutrinos and axions  
offers a new cooling mechanism for 
magnetized white dwarfs and neutron stars.  It puts a bound 
$g_{ae}\lesssim10^{-10}$ on the axion electron 
coupling constant.   
Similarly from magnetic white dwarf (WD) stars 
$g_{ae}\lesssim9\times 10^{-13}(T/10^7{\rm K})^{5/4}
(B/10^{10}{\rm G})^{-2}$, where $T$ is the internal 
temperature of the 
white dwarf.  

In this letter, we calculate the corrections to the emission rate 
of axions by nucleons in a strongly 
magnetized neutron or WD star whose magnetic field is 
stronger than the critical field, $B>B_c$, such that the Landau gap 
is larger than the rest mass energy or $\Lambda_{QCD}$. As we will see 
later, for $B\gtrsim10^{20}{\rm G}$ 
(or $\sqrt{\left|qB\right|}\gtrsim4\Lambda_{\text {QCD}}\simeq 
800~{\rm MeV}$)  
the axion emission is enhanced quite 
a lot and gives a stringent bound on the axion-nucleon coupling.  
Large magnetic fields $B\sim 10^{12}$-$10^{14}{\rm G}$ 
have been estimated  
at the surface of neutron stars from the synchrotron radiation 
spectrum~\cite{woltjer} and fields 
$B\sim 10^{18}$-$10^{20}{\rm G}$ are 
predicted in the core by scalar virial theorem~\cite{pal}. 

Let us first consider how the axion decay constant is 
renormalized in QED under a strong magnetic field. The calculation
will go in parallel in QCD also. 
Under a constant magnetic field oriented along the $z$ axis 
$\vec B=B\hat z$, 
the spectrum of charged particles of charge $q$ and mass $m$  
is quantized by the Landau level 
\begin{equation}
E_A=\alpha\sqrt{m^2+k_z^2+ 2\left|qB\right|n},
\label{energyev}
\end{equation} 
where $\alpha=\pm$ is the sign of the energy
and the quantum number $n$ labeling the Landau level
\begin{eqnarray}
2n=2n_r+1+|m_L|-{\rm sgn}(qB) (m_L+2\beta) 
\end{eqnarray}
with $\beta=\pm{1\over2}$
the spin component along the magnetic field. 
Here $n_r$ is the number of nodes of radial 
eigenfunction and $m_L$ is the
angular momentum. (Note that there will be corrections to the energy 
eigenvalue Eq.(\ref{energyev}) due to QED loop 
effects like the anomalous 
magnetic moment, but they are irrelevant since the corrections are 
smaller than the Landau gap.)

At energy lower than the Landau gap,  
$E<\sqrt{\left|qB\right|}$, the relevant degrees of 
freedom in QED are 
photons and the lowest Landau level (LLL) fermions.  Their 
interaction can be described by an effective action 
derived by integrating out the modes at the higher Landau levels 
($n>0$)~\cite{hong98}. The effective action contains a (marginal)  
four-fermion interaction, which leads to  
chiral symmetry breaking even at weak 
attraction~\cite{gusynin,ng,hong96}. 
Using this effective action, we calculate the axion decay amplitude 
to two photons. The Wilsonian effective action 
of QED is 
given as~\cite{hong98}
\begin{eqnarray}
{\cal L}_{\rm eff} & = & -{1\over4}(1+a_1)
F_{\mu\nu}^2+(1+b_1)\bar\psi \left(i\mathord{\not\mathrel{D}}
-m\right)\psi
-(1+c_1){ie^2\over 2|eB|}\bar\psi\mathord{\not\mathrel{A}}
{\tilde\gamma}_{\mu}
\stackrel{\leftrightarrow}\partial^{\mu}_{\parallel}
\mathord{\not\mathrel{A}}\psi \nonumber \\
&  & -{ie^2\over 2|eB|}
\bar\psi\mathord{\not\mathrel{A}} {\tilde\gamma}_{\mu}
\stackrel{\leftrightarrow}\partial^{\mu}_{\perp}
\mathord{\not\mathrel{A}}\psi
+{g_1\over|eB|}\left[\left(\bar\psi\psi\right)^2+
\left(\bar\psi i\gamma_5\psi\right)^2\right]+\cdots,
\label{effective1}
\end{eqnarray}
where the coefficients $a_1, b_1$, and $c_1$ are
of order $e^2$ while $g_1$ is of order $e^4$, 
determined by the one-loop matching conditions, 
and the ellipses are the higher order operators in momentum expansion.
The components $\partial_{\perp}$ ($\partial_{\parallel}$)
are perpendicular (parallel) to the 
magnetic field and
\begin{equation}
\tilde\gamma^{\mu}=
    \left\{ \begin{array}{ll}
    i\gamma^{\mu}\gamma^1\gamma^2\ln2~ {\rm sgn}(eB) &
        \mbox{if $\mu=0,3$}\\
        \gamma^{\mu} & \mbox{if $\mu=1,2$}.
    \end{array}
    \right.
\end{equation} 
Now, we consider the axion coupling to electrons:
\begin{equation}
{\cal L}_{\rm int}={1\over2f_{PQ}}\partial^{\mu}a\bar \Psi
\gamma_{\mu}\gamma_5\Psi,
\end{equation}
where $a$ is the axion field and $f_{PQ}$ is the axion decay constant.
We integrate out all the modes ($\Psi_{n\ne0}$)
except the LLL electron ($\psi$) to get  the low energy
effective interaction for $a$ and $\psi$, which is given as
\begin{equation}
{\cal L}_{\rm int}^{\rm eff}={1\over2f_{PQ}}\partial^{\mu}a\bar
\psi
\gamma_{\mu}\gamma_5\psi
-{e\over 2|eB|f_{PQ}}
\bar\psi\gamma^{\alpha}\gamma_5\partial_{\alpha}a{\tilde\gamma}^{\mu}i
\stackrel{\leftrightarrow}\partial_{\mu}
\gamma^{\beta}\psi A_{\beta}
+\cdots,
\label{effective2}
\end{equation}
where the ellipses denote the effective interaction terms containing
the higher powers of axions and momenta, generated by the exchange
interaction of $\Psi_{n\ne0}$. 

In the leading order, there are
three diagrams that contribute to the 
axion decay amplitude (see Fig. 1). 
We find that the first two diagrams vanish but the third diagram gives 
\begin{eqnarray}
A_1
&=&-{ie^2\over\left|eB\right|f_{PQ}}(2\pi)^4\delta^4(k-P)
\int_q(q+k)_{\nu}k_{\alpha}\varepsilon_{\mu}(p_1)
\varepsilon_{\beta}(p_2){\rm Tr}
\left[\gamma^{\mu}\tilde S(q+p_1)\gamma^{\alpha}\gamma_5
\tilde\gamma^{\nu}\gamma^{\beta}\tilde S(q)\right]
\\
&=&-{e^2\over 4\pi^2f_{PQ}}\delta^4(k-P)
\epsilon^{\mu\nu\alpha\beta}p_{1\mu}\varepsilon_{\nu}(p_1)
p_{2\alpha}\varepsilon_{\beta}(p_2),
\end{eqnarray}
where $P=p_1+p_2$ and $\varepsilon_{\mu}(p)$ the photon wave function,
and the LLL electron propagator
\begin{equation}
\tilde S(l)={i\over \not l_{\parallel}-m}P_-e^{-l_{\perp}^2/
\left|eB\right|}
\end{equation}
with $P_-=1-i\gamma^1\gamma^2{\rm sgn}(eB)$.
The effective Lagrangian for the axion decay amplitude is then 
\begin{equation}
{\cal L}^{\rm eff}_{a\gamma\gamma}={e^2\over 16\pi^2}{a\over f_{PQ}}
F\tilde F.
\label{axion}
\end{equation}
We find that the axion decay constant is not renormalized
in the leading order under an external magnetic field,  
$f_{PQ}(B)=f_{PQ}(0)$, 
in contrast with the pion decay constant that increases  
under an external magnetic field~\cite{smilga}. In fact, as is shown 
in the appendix A, the non-renormalization of the axion decay 
constant under external magnetic fields is exact  
simply because the axion couples to $\vec E\cdot\vec B$ 
not to magnetic fields $B$ alone. 

Similarly, we calculate the decay amplitude of the axion into
one photon (Fig. 2), using the effective Lagrangians
Eqs. (\ref{effective1}) 
and (\ref{effective2}), to get  
\begin{eqnarray}
A_2&=&{ie\over 2f_{PQ}}(2\pi)^4\delta^4(k-p)
\varepsilon_{\alpha}(p)k_{\beta}\int_q
{\rm Tr}\left[\gamma^{\alpha}\tilde S(q)\gamma^{\beta}\gamma_5
\tilde S(q-k)\right]
\\
&=&{e^2\over4\pi^2f_{PQ}}(2\pi)^4\delta^4(k-p)\epsilon^{12\alpha\beta}
\varepsilon_{\alpha}(p)k_{\beta}B.
\end{eqnarray}
This result agrees with the direct calculation done with
the effective Lagrangian Eq.(\ref{axion}), where
one of two photons is replaced by the external photon, 
assuring that $f_{PQ}$ is not renormalized. We see that 
the external magnetic field does not change the axion-photon
interaction.  Therefore, the energy-loss processes involving 
only axions and photons such as the axion-photon conversion or 
the Primakoff process do not get affected by the dimensional reduction. 

Now we consider the process that electrons or quarks emit axions.  
For an extremely strong magnetic field $B>B_c$, all the charged 
particles are in the lowest Landau level and thus the Bremsstrahlung 
emission is more relevant than the cyclotron emission, unless the 
interior temperature of stars is comparable to the Landau gap.
The one loop correction to the axion Bremsstrahlung amplitude is 
(see Fig. 3) 
\begin{eqnarray}
A_{BR}&=&{i^2g_1\over 2f_{PQ}
\left|eB\right|}\int_{x,y}
\left<p^{\prime},k\right|\bar \psi
\not\partial a\gamma_5\psi(y)
\left[\left(\bar\psi\psi(x)\right)^2+
\left(\bar\psi i\gamma_5\psi\right)^2\right]
\left|p\right>\\
&=&{-ig_1\over 4\pi^2f_{PQ}}(2\pi)^4
\delta^4(p-p^{\prime}-k)\cdot
\bar u(p^{\prime})\not\!k\gamma_5 u(p)
\left(1+{4\over3}{k^2\over m^2}+\cdots\right).
\end{eqnarray} 
We see that the Bremsstrahlung process has an extra contribution
coming from the effective four-fermion interaction.
(Note that here we assume $k^2\ll m^2$. But, if $m\to0$,
the contribution goes to zero.)
In QED, $g_1\propto e^4$ and therefore quite small~\cite{hong98}, but 
in QCD, for a strong magnetic field $B>10^{19}~G$ (or 
$\sqrt{|qB|}>\Lambda_{\text {QCD}}$) such that  
the effective action for QCD  
is valid,  $g^s_1\gtrsim \alpha_s^2$ and 
is no longer negligible 
(see the appendix B for details). Especially, for quarks under a strong 
external magnetic field $B\gtrsim10^{20}~G$, $g^s_1$ will 
be quite large.
Thus the four-quark interaction will be stronger than the one-gluon 
exchange interaction and will break the chiral symmetry of QCD   
at energy above $\Lambda_{\text {QCD}}$. 

The axion emission process of neutron stars was first calculated
by Iwamoto and it is found that
axion bremsstrahlung from nucleon-nucleon collision
is the dominant energy-loss process in neutron stars~\cite{iwamoto}.
Since the emission rate is proportional to
the square of the axion-nucleon coupling $g_{an}^2$,
any correction to this coupling will affect the emission rate.
What we calculated above can be thought of as a finite correction
to the axial singlet current of quarks and equally valid to the axial
octet quark currents. Namely, 
both $\bar q\gamma_{\mu}T^a\gamma_5q$ and
$\bar q\gamma_{\mu}\gamma_5q$ have the same correction
$(1+g^s_1/(2\pi^2))$.
As was discussed in Ref.~\cite{kaplan}, the derivative couplings
of the axion to the axial vector baryon current is
\begin{equation}
{\partial^{\mu}a\over f_{PQ}}\left\{2{\rm tr}\left(Q_AT^a\right)
\left(F{\rm tr}\bar B\gamma^{\mu}\gamma_5\left[T^a,B\right]
+D{\rm tr}\bar B\gamma^{\mu}\gamma_5\left\{T^a,B\right\}
\right)
+{S\over3}{\rm tr}Q_A{\rm tr}\bar B\gamma^{\mu}\gamma_5 B
\right\},
\end{equation}
where $B$ is the baryon octet matrices and $Q_A$ the axion charge.
According to the current algebra, $F$, $D$, and $S$ are defined
to be proportional to the nucleon matrix elements of the axial
quark currents~\cite{manohar}.
Therefore we see that the axion-nucleon
coupling gets a correction
\begin{equation}
\delta g_{an}= {g^s_1\over2\pi^2}g_{an}.
\end{equation}

Besides the correction due to the four-fermion interaction, 
the emission rate gets corrections since the axion-nucleon coupling and 
the nucleon mass will change due to the strong magnetic field. 
As a good approximation, we can take the nucleon mass as  
\begin{equation}
m_n\propto\left<\bar qq\right>^{1/3}.
\end{equation}
In chiral perturbation theory, the quark condensate is  
given as~\cite{leutwyler}
\begin{equation}
\left<\bar qq\right>^{B}=-{\partial {\cal E}_{\text {vac}}(m_q, B)
\over \partial m_q}\Big|_{m_q=0},
\end{equation}
where ${\cal E}_{\text{vac}}$ is the vacuum energy density. 
The one-loop vacuum energy under 
a magnetic field was calculated by Schwinger~\cite{schwinger}   
as 
\begin{equation}
{\cal E}_{\text {vac}}(B)={\cal E}_{\text vac}(0)
-{1\over16\pi^2}\int_0^{\infty}{ds\over s^3}e^{-M_{\pi}^2s}
\left[{eBs\over \sinh (eBs)}-1\right].
\end{equation}
Therefore, the condensate under an external magnetic field is,  
using the Gell-Mann-Oakes-Renner relation
$F_{\pi}^2M_{\pi}^2=\left<\bar qq\right>(m_u+m_d)$,   
\begin{equation}
\left<\bar qq\right>^{B}=\left<\bar qq\right>^{B=0}
\left(1+{\left|eB\right|\ln2\over 16\pi^2F_{\pi}^2}+\cdots\right),
\end{equation}
where the ellipses denote the higher loop corrections~\cite{smilga}. 
Since the axion-nucleon coupling is proportional to the nucleon mass 
($g_{an}\propto m_n/f_{PQ}$) 
and the axion decay constant does not change under magnetic fields,
\begin{equation}
g_{an}(B)=g_{an}(0)
\left(1+{\left|eB\right|\ln2\over 48\pi^2F_{\pi}^2}+\cdots\right).
\end{equation} 
The energy-loss rate per unit volume due to 
the nucleon-nucleon axion bremsstrahlung is calculated in 
Ref.~\cite{iwamoto,turner,ellis} and can be written as 
\begin{equation}
{\cal Q}^1_a\propto \left(f\over M_{\pi}\right)^4
m_n^{2.5}g_{an}^2T^{6.5},
\end{equation}
where $f$ is the pion-nucleon coupling and $T$ is the temperature. 
Therefore, we find the correction  
due to the strong magnetic field is 
\begin{equation}
{\cal Q}_a^1(B)={\cal Q}_a^1(0)
\left(1+{g^s_1\over2\pi^2}+\cdots\right)^2
\left(1+{\left|eB\right|\ln2\over 16\pi^2F_{\pi}^2}+
\cdots\right)^{2.5/3},
\end{equation}
where we have used the Goldberg-Treiman relation $m_n=f F_{\pi}$ 
together with the Gell-Mann-Oakes-Renner relation. 
Similarly, the lowest-order energy emission rate per unit volume by the 
pion-axion conversion $\pi^-+p\to n+a$~\cite{schramm} 
has corrections due to the strong magnetic field as 
\begin{equation}
{\cal Q}_a^{\pi^-}(B)={\cal Q}_a^{\pi^-}(0)
\left(1+{g^s_1\over2\pi^2}+\cdots\right)^2
\left(1+{\left|eB\right|\ln2\over 16\pi^2F_{\pi}^2}+
\cdots\right)^{-1/3},
\end{equation}
Note that the above derivation of the change in the quark condensate 
relies on the chiral perturbation theory 
and the one loop vacuum energy calculation, which will not be valid  
presumably for an extremely strong magnetic field such 
that the Landau gap  
becomes bigger than the $\rho$ meson mass 
$\sqrt{\left|eB\right|}>m_{\rho}$.
But, the condensate will still increase as $B$ even for such a strong  
magnetic field because the increment of vacuum energy due to the 
external field will be balanced by forming larger condensates.

The correction to the emission rate due to the external magnetic field 
comes from two sources. First one is from the finite correction to the 
axial current due to the effective four-quark operator 
and the second one is 
by the change of the quark condensate under the external field. 
For $B=10^{18}\sim10^{19}~G$, the correction by the second one 
is less than a few percent while the correction by the first one is 
of order one. 
For $B=10^{20}~G$ or $\sqrt{\left|eB\right|}=0.77 {\text {GeV}}$, 
the correction by the second is about 24\% for the Bremsstrahlung and 
about  $-10\%$ for the pion-axion conversion,  
while the correction 
due to the first one is about $0.06\alpha_s$ for both processes, 
because the four-quark operator will evolve sufficiently. 
For $B\gtrsim 10^{20}~G$  
the perturbative calculation breaks down and we expect that 
the correction will be extremely large because the strong 
enhancement of $g_1^s$ at low energy.  
Therefore, if $B\gtrsim 10^{20}~G$ in stars, they will emit 
axions too quickly and will not survive present.  On the other hand, if 
we observe such a strongly magnetized star, the axion-nucleon 
coupling has to be smaller than the current upper bound by 
several orders of magnitude.

In conclusion, we have shown that the axion decay constant 
does not get a correction by the external magnetic field and we 
have calculated the correction to the axion-nucleon 
coupling due to external magnetic fields. We find that if 
$B\gtrsim 10^{20}~G$ in strongly magnetized neutron 
and white dwarf stars, the  
axion emission rate is enhanced by  
several orders of magnitude.

\acknowledgments

The author is grateful to S. Chang, John Ellis, 
Pierre Ramond, and Pierre 
Sikivie for discussions and also to the members of the IFT of 
University of Florida for their hospitality during his stay at IFT. 
This work was supported in part by the KOSEF through SRC program of
SNU-CTP and also by the academic research fund of Ministry of 
Education, Republic of Korea, Project No. BSRI-97-2413. 
The author wishes to acknowledge the financial support of 
Korea Research Foundation made in the program year of 1997.

\appendix
\section{QED vacuum energy}

The low energy effective action of QED, integrating out 
fermions and scalars, has been calculated for various 
external fields~\cite{heisenberg}. 
Here, we calculate the QED vacuum energy, 
integrating out the axion fields:
\begin{eqnarray}
e^{-\int_x {\cal E}_{\text{vac}}(E,B)}
=\int d[a] \exp\left[-\int_x \left({1\over4}F^2-
{1\over2}\left(\partial_{\mu}a\right)^2+
V_0\cos \left({a\over f_{PQ}}\right)
+{ie^2\over 16\pi^2}{aF\tilde F\over f_{PQ}}\right)\right],
\label{energy}
\end{eqnarray}
where we neglect other matter fields for simplicity but 
it does not change our argument.
The vacuum energy density is then
\begin{equation}
{\cal E}_{\text{vac}}(E,B)={1\over2}(E^2+B^2)+
{\cal E}^a_{\text{vac}}(E, B),
\end{equation}
where the second term is due to axions and obviously a function 
of $F\tilde F$ or $\vec E\cdot\vec B$. 
If $\vec E\cdot\vec B=0$, the vacuum energy does not change and 
neither does the axion decay constant. 
But, if $\vec E\cdot\vec B\ne0$, the vacuum energy increases by 
the axion loop contribution because the axion-photon coupling in 
Euclidean space is imaginary. Using Schwarz inequality, one 
can easily show that ${\cal E}^a_{\text{vac}}(E, B)\ge0$.   
Thus under the external fields the condensate has to  
increase to balance the vacuum energy and 
the axion decay constant gets bigger.    
For general electromagnetic fields,  we get   
$f_{PQ}(E,B)=f_{PQ}(\vec E\cdot\vec B)\ge f_{PQ}(0)$.

If we keep only the axion mass term 
in the potential in Eq. (\ref{energy})  
and integrate over $a$, we get for slowly varying fields 
\begin{equation}
{\cal E}^a_{\text{vac}}(E, B)\simeq
8\left({e^2\over 16\pi^2} \right)^2 
\left({\vec E\cdot \vec B\over \Lambda^2_{\text{QCD}}}\right)^2, 
\end{equation}
where we used the axion mass $m_a=\Lambda^2_{\text{QCD}}/f_{PQ}$.

\section{QCD under a constant external magnetic field}

Since quarks have electric charges, magnetic catalysis will operate for
quarks as well when the external magnetic field is sufficiently strong, 
$B>10^{19}~G$ or $\sqrt{|qB|}>\Lambda_{\text {QCD}}\simeq 200{\rm MeV}$.
By the renormalization group analysis, we will see that if 
the magnetic field is stronger than
$10^{20}~G$, the four-quark interaction is stronger than the one-gluon 
exchange interaction and thus  
the chiral symmetry will break at scale higher than
$\Lambda_{\text {QCD}}$.
The derivation of low energy effective Lagrangian will be same as that
of QED except the group theoretical factors.
At low (but quite higher than $\Lambda_{\rm QCD}$)
energy the relevant degrees of freedom are LLL quarks, gluons,
and photons. But, since the gluons couple to quarks more strongly than
photons, we will
neglect the photon interactions. As in QED,
the exchange of quarks at the higher Landau level will generate a
tree-level interaction of quarks and gluons, which will induce
a four-quark interaction at one-loop matching (see Fig. 4);
\begin{equation}
{\cal L}^1_{\rm eff}\ni {g_1^{s}\over 4\left|qB\right|}\left[\left(
\bar Q_0 Q_0\right)^2+\left(\bar Q_0 i\gamma_5 Q_0\right)^2\right],
\end{equation}
where $Q_0$ is the LLL quark and $q$ is the electric charge of quark.
The one-loop RG equation for the four-quark interaction is given as
\begin{equation}
\mu{d\over d\mu}g_1^s=-{40\over9}\alpha_s^2(\ln2)^2,
\label{rg1}
\end{equation}
where $\alpha_s$ is the coupling constant of strong interaction.
Unlike QED, the strong coupling constant runs at scale below the
Landau gap $\Lambda_L=\sqrt{\left|qB\right|}$ as~\cite{ramond}
\begin{equation}
{1\over \alpha_s(\mu)}={1\over \alpha_s(\Lambda_L)}+{11\over2\pi}
\ln{\mu\over \Lambda_L}.
\label{rg2}
\end{equation}
Note that quarks do not contribute to the running coupling since
the quark loop is finite due to the dimensional reduction.
Combining the above RG equations (\ref{rg1}) and (\ref{rg2}),
we find how the four-quark coupling changes as
scale;
\begin{equation}
g_1^s({\mu})=1.1424 \left(\alpha_s(\mu)-\alpha_s(\Lambda_L)\right)
+g_1^s(\Lambda_L),
\end{equation}
where $g_1^s(\Lambda_L)$ is of the order of $\alpha_s(\Lambda_L)^2$.
We see that the four-quark coupling becomes stronger than the
strong coupling constant at a scale
larger than the QCD scale $\mu>\Lambda_{\text {QCD}}$ if  
$\Lambda_L\gtrsim4 \Lambda_{\text {QCD}}$ or 
$B\gtrsim 10^{20}~G$. For such a strong magnetic 
field, the chiral symmetry of QCD will be broken at scale larger 
than $\Lambda_{\text {QCD}}$,
the usual chiral symmetry breaking scale of QCD.

\newpage 
\begin{figure}
\centerline{\epsfbox{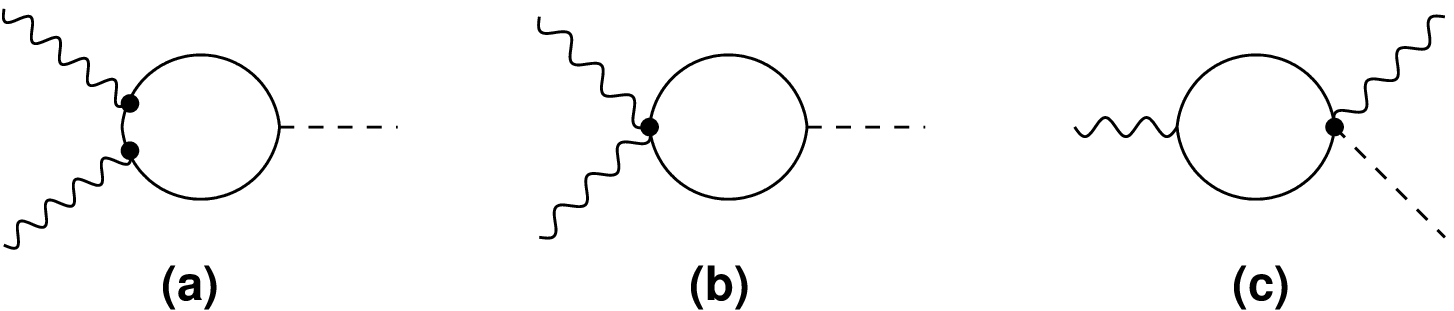}}
\caption{Axion decay amplitude. Wiggly lines denote photons, 
solid lines the LLL fermions, and broken lines axions. }
\end{figure}
\begin{figure}
\centerline{\epsfbox{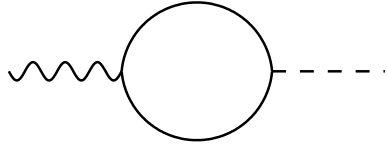}}

\caption{Axion conversion into a photon. Solid line denote 
electrons.}
\end{figure}
\begin{figure}
\centerline{\epsfbox{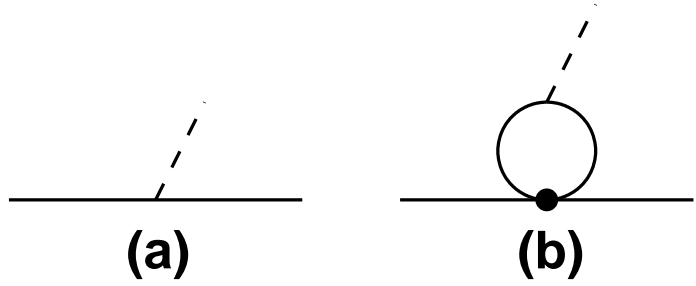}}

\caption{Axion Bremsstrahlung emission. Solid lines denote electrons 
or quarks}
\end{figure}
\begin{figure}
\centerline{\epsfbox{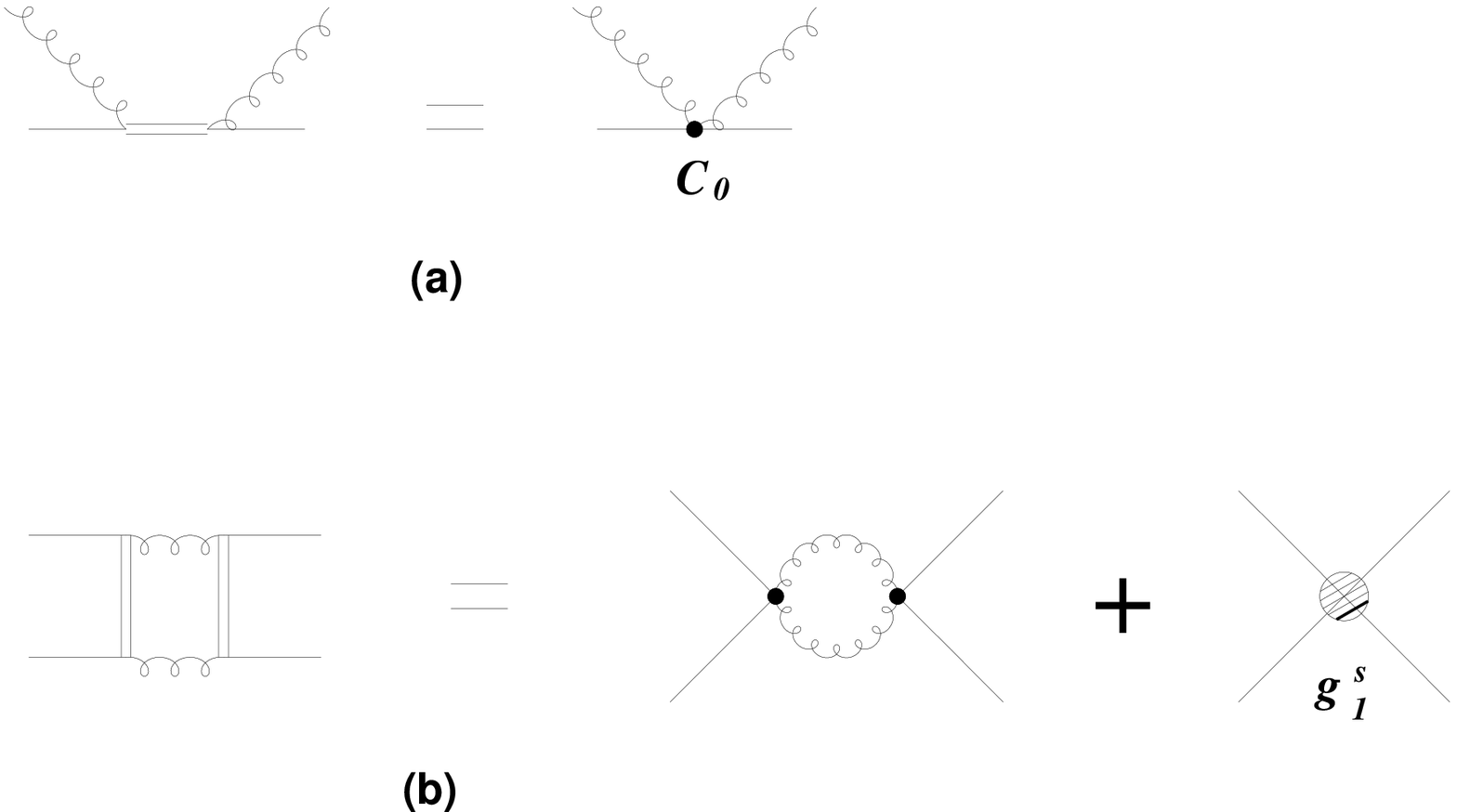}}

\caption{Matching conditions. Double lines denote quarks at 
the higher Landau levels ($n>0$), single line LLL quarks, and 
curled lines gluons. (a) tree level, (b) one loop level}
\end{figure}


\begin{references}

\bibitem{raffeltreport}For a review, see M. S. Turner, Phys. Rep. 
{\bf 197}, 67 (1990); G. Raffelt, {\it ibid.}  
{\bf 198}, 1 (1990); 
{\it Stars as Laboratories for Fundamental Physics}
(University of Chicago Press: Chicago 1996). 
\bibitem{hagman}For the axion search in laboratories, see 
C. Hagman {\it et al.}, Nucl. Phys. Proc. Suppl. {\bf 51B}, 209 (1996);
I. Ogawa, S. Matsuki, and K. Yamamoto, Phys. Rev. D {\bf 53}, R1740 
(1996).  
\bibitem{seckel}G. Raffelt and D. Seckel, Phys. Rev. Lett. 
{\bf 67}, 2605 (1991); Phys. Rev. D {\bf 52}, 1780 (1995);
H.-Th. Janka, W. Keil, G. Raffelt, and D. Seckel, 
Phys. Rev. Lett. {\bf 76}, 2621 (1996).  
\bibitem{borisov}A. V. Borisov and V. Yu. Grishina, JEPT {\bf 79},
837 (1994).
\bibitem{kachelriess}M. Kachelrie\ss, C. Wilke, 
and G. Wunner, Phys. Rev.
D {\bf 56}, 1313 (1997).
\bibitem{schramm}R. Mayle,
D. N. Schramm, M. S. Turner, and J. R. Wilson, Phys. Lett. B
{\bf 317}, 119 (1993); M. S. Turner, Phys. Rev. D {\bf 45}, 1066
(1992).
\bibitem{ellis}W. Keil {\it et al.}, 
Phys. Rev. D {\bf 56}, 2419 (1997).
\bibitem{woltjer}C. Woltjer, Astrophys. J. {\bf 140}, 1309 (1964);
T.A. Mihara {\it et al.}, Nature {\bf 346}, 250 (1990); G. Ghanmugan, 
Annu. Rev. Astron. Astrophys. {\bf 30}, 143 (1992).
\bibitem{pal}S. L. Shapiro and S. A. Teukolsky, {\it Black Holes, 
White Dwarfs and Neutron Stars} (Willey, New York, 1993);
S. Chakrabarty, D. Bandyopadhyay, and S. Pal, 
Phys. Rev. Lett. {\bf 78}, 2898 (1997); D. Bandyopadhyay, 
S. Chakrabarty, 
and S. Pal,
{\it ibid.} {\bf 79}, 2176 (1997).
\bibitem{hong98}D.K. Hong, hep-ph/9707432, to appear in Phys. Rev. D,
March 15, 1998.
\bibitem{gusynin}V.P. Gusynin, V.A. Miransky, and I.A. Shovkovy,
Phys. Rev. Lett. {\bf 73} (1994) 3499; Phys. Rev. D {\bf 52}, 4747
(1995); Nucl. Phys. B {\bf 462}, 249 (1996).
\bibitem{ng}C.N. Leung, Y. J. Ng and A.W. Ackley,
 Phys. Rev. D {\bf 54}, 4181 (1996); D. S. Lee, C. N. Leung, and 
Y. J. Ng, {\it ibid.} {\bf 55}, 6504 (1997).
\bibitem{hong96}D. K. Hong, Y. Kim, and S. Sin, Phys. Rev. D {\bf 54},
7879 (1996).
\bibitem{smilga}I. A. Shushpanov and A. V. Smilga, 
Phys. Lett. B {\bf 402}, 351 (1997). 

\bibitem{iwamoto}N. Iwamoto, Phys. Rev. Lett. {\bf 53}, 1198 (1984).
\bibitem{kaplan}D. Kaplan, Nucl. Phys. {\bf B260}, 215 (1985);
M. Srednicki, {\it ibid.} {\bf B260}, 689 (1985);
H. Georgi, D. Kaplan, and L. Randall, Phys. Lett. {\bf B169}, 73
(1986).
\bibitem{manohar}R. Shrok and L.-L. Wang, Phys. Rev. Lett. {\bf 41},
1692 (1978);R. Jaffe and A. Manohar, Nucl. Phys. {\bf B337}, 509 (1990).
\bibitem{leutwyler}J. Gasser and H. Leutwyler, Nucl. Phys. 
B {\bf 250}, 465 (1985).
\bibitem{schwinger}J. Schwinger, Phys. Rev. {\bf 82}, 664 (1951);
{\it Particles, Sources, and Fields} (Addison-Wesley, 1973).

\bibitem{turner}R. P. Brinkmann and M. S. Turner, Phys. Rev. D {\bf 38},
2338 (1988).
\bibitem{ramond}H. Arason {\it et al.} 
Phys. Rev. D {\bf 46}, 3945 (1992).
\bibitem{heisenberg}See, for example, J. S. Heyl and L. Hernquist, 
Phys. Rev. D {\bf 55}, 2449 (1997).

\end{references}
\end{document}